# Surrogate cloud fields with measured cloud properties

Victor Venema, Susanne Crewell, Clemens Simmer

E-mail: Victor.Venema@uni-bonn.de, phone: +49 (0)228 735185, fax: +49 (0)228 735188

Meteorological Institute Bonn University, Auf dem Hügel 20, 53121 Bonn, Germany

## ABSTRACT

This paper describes two new methods to generate 2D and 3D cloud fields based on 1D and 2D ground based profiler measurements. These cloud fields share desired statistical properties with real cloud fields. As they, however, are similar but not the same as real clouds, we call them surrogate clouds. One important advantage of the new methods is that the amplitude distribution of cloud liquid water is also exactly determined by the measurement: The surrogate clouds made with the classical methods such as the Fourier method and the Bounded Cascade method are Gaussian and 'log-normal-like', respectively. Our first new method iteratively creates a time series with a measured amplitude distribution and power spectrum. Our second method uses an evolutionary search algorithm to generate cloud fields with practically arbitrary constraints. These clouds will be used to study the relation between radiation and cloud structure.

## INTRODUCTION

To perform 3D-radiative transfer calculations we need 3-dimensional cloud fields. However, measurements only provide 1D or 2D statistics. The typical solution to such a problem is to use the measurements to validate a (cloud) model, and use the model output for the (radiative transfer) calculations. However, it is extremely difficult to obtain cloud fields that have spe-cific properties using a cloud model. One can already be satisfied if the model produces a cloud with about the same average Liquid Water Path (LWP) and average cloud height as the ones which have been observed.

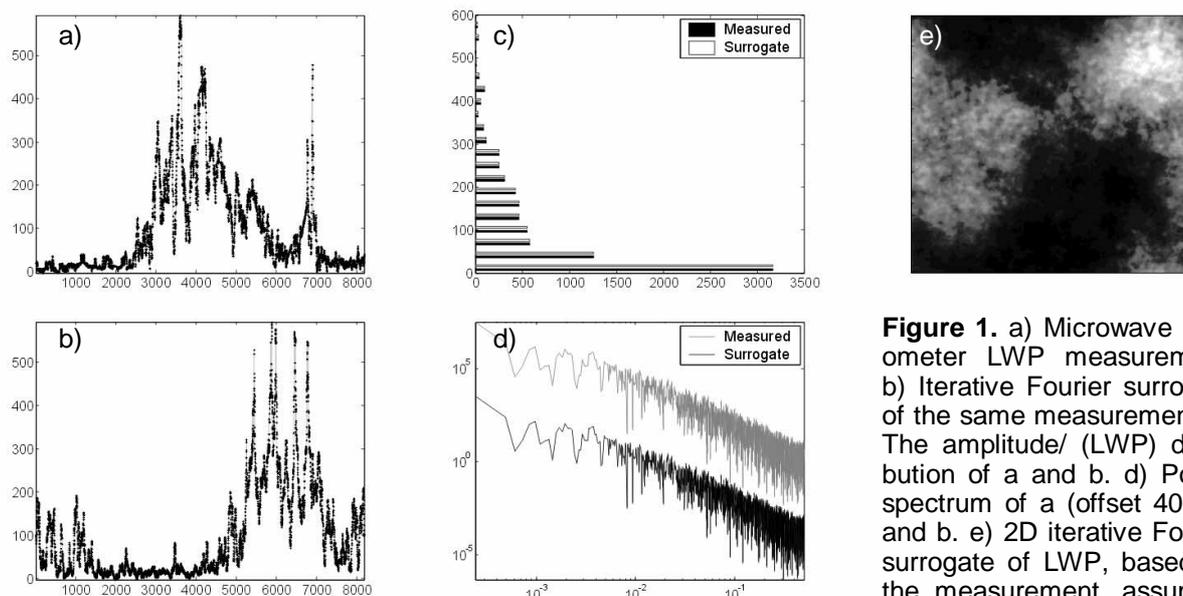

**Figure 1.** a) Microwave radiometer LWP measurement. b) Iterative Fourier surrogate of the same measurement. c) The amplitude/ (LWP) distribution of a and b. d) Power spectrum of a (offset 40 dB) and b. e) 2D iterative Fourier surrogate of LWP, based on the measurement, assuming isotropy.

In this paper, we will introduce two methods that are able to produce surrogate cloud fields out of 'sparse' measurements without using a model. These fields correspond closely to the measurements. However, you do need a statistical description that is appropriate to your question. Our goal is to generate surrogate cloud fields that are based on combined radar-lidar boundary statistics, microwave radiometer cloud water data, and IR radiometer cloud cover information.

## ITERATIVE SURROGATES

Iterative Fourier surrogates were developed by Schreiber and Schmitz (1996, 2000) for a totally different purpose for use in the non-linear time series analysis. This method is able to produce cloud fields with the amplitude distribution and power spectrum of a measured time series. From a measured time series, $s_n$ (with n time or space), with N values, the power spectrum ($S_k^2$; with k wave number) is calculated,

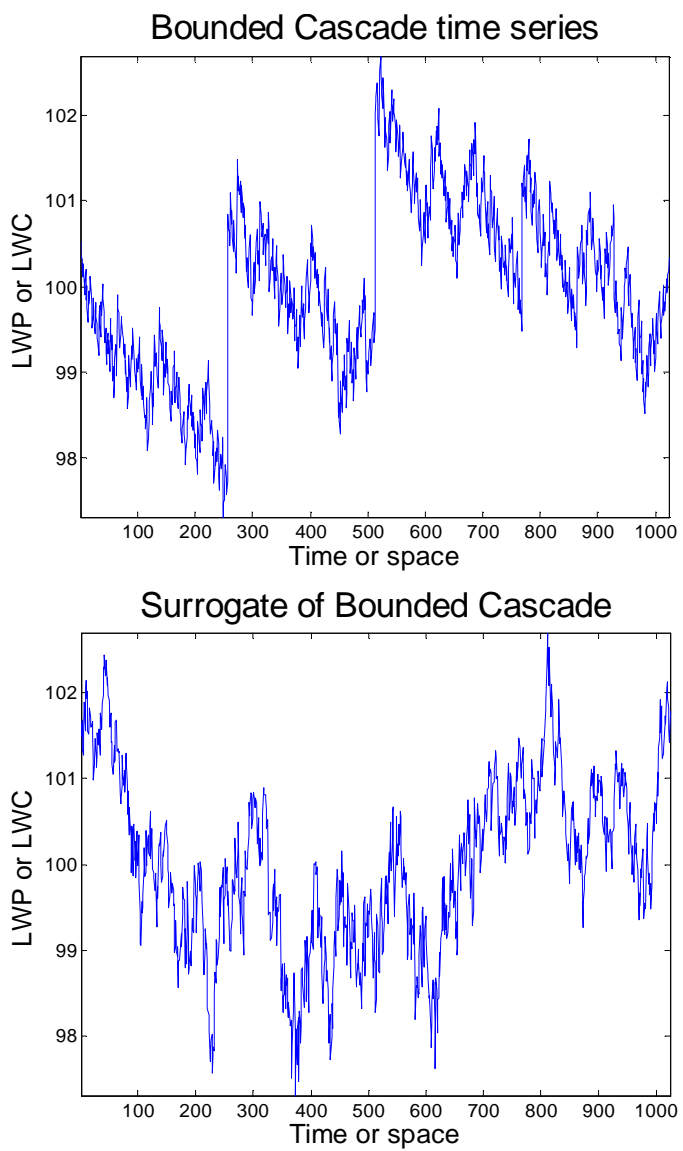

**Figure 2.** a) A surrogate time series made with the non-linear Bounded Cascade (BC) method. b) An iterative Fourier surrogate with the same amplitude distribution and power spectrum.

$$S_k^2 = \left| \sum_n s_n e^{i 2\pi k n / N} \right|^2 \qquad (1),$$

and a sorted list is stored of the values of $s_n$. The algorithm starts with a random shuffle of the data points. Then an iterative process starts in which first the Fourier coefficients are adjusted and second the amplitudes. To achieve the desired power spectrum the Fourier transform of the iterate time series is calculated and its squared coefficients are replace by those from the original time series. The phases are kept unaltered. After this step the amplitudes will no longer be the same. Therefore in the second step the amplitudes are adjusted by sorting its values and replacing them by the original sorted list of all values. This procedure has to be repeated a few times until the quality is sufficient. The 1D surrogate (fig. 1) shows that the convergence is close to perfect. The 2D surrogate is the wanted end-product, in this case. The 2D-Fourier-coefficients were calculated assuming isotropy: If one integrates over all 2D coefficients with $|k| = \sqrt{k_1^2 + k_2^2}$ one gets the 1D-coefficient for k. A next step would be to generate the full 3D LWC fields using LWC-profile time series (Löhnert and Crewell, 2002).

Schreiber and Schmitz (2000) show that it is also possible to iterate two time series simultaneously (additionally taking into account their cross power spectrum). This could be interesting to make cloud fields using LWP and cloud top statistics simultaneously; including cloud top is important for cloud remote sensing applications. Maybe it could even work to include cloud base information as a third time series and then combine these three 2D fields with the measured average LWC profile to get a 3D surrogate cloud field.

This Fourier method does have its limitations. Figure 2a shows a time series made with the Bounded Cascade (BC) method (Cahalan, 1994) and an iterative Fourier surrogate (2b) with the same amplitude distribution and power spectrum. One can see that the two structures are similar, but not fully the same. The original BC time series shows large jumps, which are typical; these jumps are not present in its surrogate. This illustrates that the amplitude distribution and the power spectrum do not fully describe structure in case the time series was "made" with non-linear methods. As the dynamics of cloud is known to be non-linear, additional measures (next to the power spectrum or the autocorrelation function) may be necessary to describe their structure.

## SEARCH SURROGATES

The most general way to generate a cloud field is to *search* for a field that has the wanted cloud properties out of the set of all possible fields. Such a surrogate cloud field can have any (statistical) property. For this paper we made surrogate clouds using one point and two point statistics of LWP, LWC, cloud top and base height and the number of layers. As search algorithm we have used an evolutionary algorithm (Mitchell, 1996, Goldberg, 1989). Due to the badly behaved cost function, you need a very robust global search algorithm.

The idea and vocabulary of evolutionary algorithms derives from the natural evolution of species. The most important difference from other search methods is that evolutionary algorithms work on a set (population) of solutions, in our case 40 to 200 clouds. From each of these clouds the fitness function (similar to a cost function) is calculated, and the best clouds have a better chance to reproduce. During reproduction the clouds are mutated to introduce new diversity.

To calculate the overall fitness of a cloud, the fitness functions of the various cloud properties have to be combined into one number. We do this by normalizing the fitness functions to zero median and unit maximum prior to combining them to one overall fitness value. This normalization is chosen as the fitness functions often have strong negative outliers.

At the moment we are still testing the method. As test input we use statistics derived from a cloud measured with the Delft Atmospheric Research Radar. Within these cloud boundaries a linear LWC profile was assumed. From this cloud we are making surrogates that correspond to the measurements with regard to: the histograms of cloud top and base height, LWP, LWC and the number of cloud layers, furthermore, the power spectra of cloud top and base height, LWP and number of layers, as well as (as second measure of structure) the length of the cloud top and base height, LWP and LWC (vertical and horizontal). The length is defined as the sum of the absolute difference between all neighbor pixels or columns. The noisy power spectrum has to be lightly smoothed to get convergence. To calculate the cloud (64x64 pixels) in figure 3, took 22 hours on a desktop PC, with the search algorithm written in IDL. The surrogate corresponds very well to all the mentioned statistical parameters, except for the structure measures for the number of cloud layers (its power spectrum and length). To make a three dimensional cloud is a trivial, but calculation intensive extension. We are working on a faster parallelized C++ version to make this step. As the number of clouds that the algorithm has to calculate is about a linear function of the number of pixels, it is relatively unproblematic to make larger cloud fields.

## CONCLUSIONS AND OUTLOOK

The iterative Fourier method is able to generate realistic looking cloud fields. In the future, we want to find ways to get real-istic 3D-LWC fields with this method (to study radiative fluxes) and to make multiple time series (with their cross-spectra), e.g. LWP and cloud top height. This iterative method does have its limitations. Such a method based on Fourier spectra is not able to handle zero or more than one value at a certain time. Therefore, the cloud top height time series can not have a gap in the cloud or two cloudy parts in one column. A three dimensional LWC field will not necessarily produce the right cloud boundary (e.g. cloud top) statistics. Furthermore, in power spectrum may not be all the information about cloud structure that we need for radiative transfer problems.

The search algorithm is a much more general method to generate surrogate cloud fields. It can handle (almost?) any statistical or non-statistical cloud property, at the price of being very calculation intensive. An interesting non-statistical cloud property we would like to try is, e.g., to search for a cloud field where the clouds are located at the same position as indicated by an airborne imager. In stead of the noisy power spectrum we would like to use the autocorrelation function in future, which could handle cloud gaps. Alternatively, one could use the coefficients of the box counting method (normally used to calculate the fractal dimension) to describe the structure of the cloud boundaries.

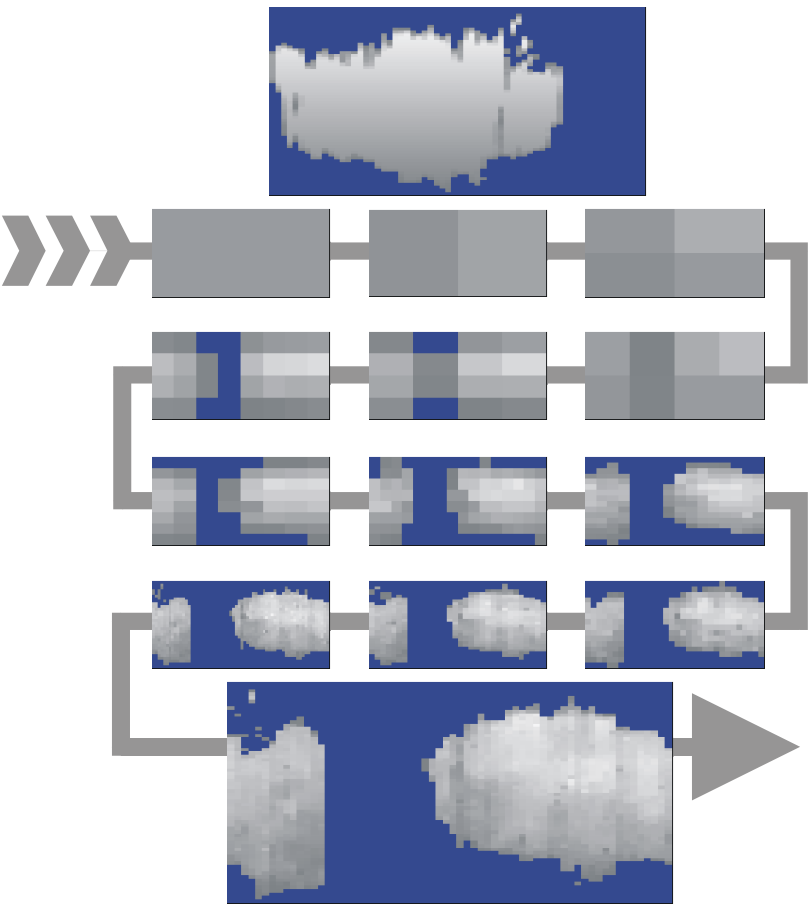

**Figure 3.** The cloud at the top was used to derive the statistics for the cloud shown below. The search starts at low resolution and following the arrow, increases its resolution each step by a factor of 2.

To get an idea of the quality of both kinds of surrogate clouds, we want to take a set of template clouds (from an LES model) and make a surrogate cloud from each template. The difference in the radiative properties between the cloud pairs indicates the accuracy of surrogates with these statistical properties.